%% file: CC_proceedings.tex
\documentclass[a4paper]{jpconf}
\usepackage{graphicx}
\usepackage{amsmath}
\usepackage{amsfonts} 
\usepackage{amssymb}
\usepackage{bbm} 
\usepackage{times}

\include{texdefs}

\begin{document}
\title{
Nonexistence of locally but not globally supersymmetric orbifolds 
}

\author{Stefan Groot Nibbelink}

\address{
Institute of Engineering and Applied Sciences, 
Rotterdam University of Applied Sciences, \\ 
G.J.\ de Jonghweg 4 - 6, 
3015 GG Rotterdam, 
The Netherlands
}

\ead{s.groot.nibbelink@hr.nl}

\begin{abstract}
Motivated by the smallness of the cosmological constant we investigate whether it is possible to have vanishing one-loop heterotic string partition functions for six-dimensional non-supersymmetric toroidal orbifolds. 
A straightforward way to realize this presents itself, when each orbifold sector separately preserves some Killing spinors, but none of them survives in all sectors combined. 
By applying some representation theory to the abstract finite point groups underlying toroidal orbifolds it turns out, that this is never possible. 
This leads to a nonexistence proof of locally but not globally supersymmetric orbifolds.  
\end{abstract}

\section{Motivation}

The current status of particle physics and cosmology is one of great success in light of the experimental confirmations of their Standard Models and confusion about what could be next. The link of the Standard Model of Particle Physics that had been elusive for a long time, the Higgs particle, is now being measured with ever increasing precision. Similarly, the Standard Model of Cosmology   is being confirmed with ever higher experimental scrutiny. One of its ingredients is that the universe is equipped with a tiny, yet, non-vanishing cosmological constant $\gL$: It is $10^{-120}$ orders of magnitude smaller than its natural gravitational scale. Any attempt to understand this vacuum energy within the realm of field theory has failed so far.

String theory comes with the promise of being a unified quantum theory of all interactions. In particular, gravity, gauge interactions and chiral particle spectra arise at the same stage in the construction of the heterotic string. Even though there are various theoretical arguments why nature at its most fundamental level should be supersymmetric and string theories are best understood with some amount of target space supersymmetry present, there has been absolutely no experimental evidence for supersymmetry so far. Especially the LHC experiments are now probing well into the region where by naive naturalness and hierarchy arguments one would have expected the first, if not many, hints for supersymmetry. Moreover, a positive cosmological constant seems to be inconsistent with supersymmetry. Hence, if string theory really describes our universe, the smallness of the vacuum energy should be investigated in the context of non-supersymmetric string theory, which is much harder to access and control theoretically.

One of the main selling points of string theory is that, contrary to generic quantum field theories, string theory is free of divergences.  Even the cosmological constant is a finite and calculable quantity. Given that all string theories are presumably related by dualities, in principle any question can be investigated within various string constructions. In particular, perturbative heterotic string theory is a well-studied subject, hence it is clear how the vacuum energy can be computed. 

The fact, that the cosmological constant is so tiny, may be taken as an indication that the cosmological constant should vanish perturbatively to all orders and only arises due to non-perturbative effects. For this to be feasible the cosmological constant should vanish at least at the one-loop level. In string theory this corresponds to the well-studied torus partition function: When it is integrated over the fundamental domain of the torus, the one-loop  vacuum energy is determined. One-loop vanishing cosmological constants have been obtained for non-supersymmetric asymmetric orientifold constructions \cite{Kachru:1998hd,Blumenhagen:1998uf,Satoh:2015nlc}. 

However, such models focussed on the gravitational sector alone, the real challenge is to have a very tiny cosmological constant and at the same time to realize the Standard Model particles. 
Non-supersymmetric heterotic strings \cite{Dixon:1986iz,AlvarezGaume:1986jb}
have been investigated in the past \cite{Nair:1986zn,Ginsparg:1986wr,Taylor:1987uv,Dienes:1994np,Sasada:1995wq,Font:2002pq,Dienes:2006ut}
and have gained renewed interest \cite{Blaszczyk:2014qoa,Angelantonj:2014dia,Faraggi:2014eoa,Abel:2015oxa,Florakis:2016ani}
in light of the elusiveness of supersymmetry in accelerator experiments. 
The construction of non-supersymmetric heterotic models with spectra that are very close to that of the Standard Model (not its supersymmetric extension) has been quite successful \cite{Faraggi:2007tj,Ashfaque:2015vta,Blaszczyk:2015zta,Nibbelink:2015vha}.

If one starts to think about generic non-supersymmetric string constructions one quickly realizes that this is a daunting task.  
Since string theories tend to live in ten dimensions, a compactification of six dimensions is mandatory. As there is an infinite (over-countable) number of six dimensional manifolds this becomes a very difficult endeavour. 
Fortunately, the number of six dimensional toroidal orbifolds is finite and strings can be exactly quantized on them. Hence, the close to 29 million such orbifolds provide a large but trackable testing ground for such investigations. 

In this work we aim to achieve the vanishing of the one-loop vacuum energy by having the integrant, the one-loop partition function to vanish. We will show that the only orbifolds for which this is possible are in fact the known supersymmetric ones. This show that also in string theory obtaining a systematic solution to the cosmological constant problem in a non-supersymmetric setting is very challenging.

\section*{Overview}

In this proceedings, based on \cite{GrootNibbelink:2017luf}, we first recall some necessary aspects of the space group description of toroidal orbifolds. After that we determine the conditions for the one-loop partition function to vanish\footnote{
Using the contributions of the non-level matched proto-graviton and proto-gravitrino states in the partition function one can argue, that in non-supersymmetric heterotic string constructions it is never possible to have the partition function to vanish \cite{Abel:2015oxa,Dienes:1990ij}. 
It is interesting to see how this conclusion is confirmed by a rather involved analysis of toroidal orbifolds. 
}. This helps us to characterize candidate non-supersymmetric orbifold geometries. Next we make use of the classification of toroidal orbifolds to identify the candidate geometries. Using those explicitly or alternatively making use of abstract finite group theory, we show that no such non-supersymmetric orbifolds exist.

\section{Space group description of toroidal orbifolds}

The space group provides a convenient language to describe toroidal orbifolds \cite{Dixon:1985jw,Dixon:1986jc}. 
A toroidal orbifold $T^6/\mathbf{G}$ can be build as follows. The starting point is a six dimensional lattice spanned by six basis vectors $e_i$: 
\equ{
\gG ~=~  \{ e\, m  | m \in \Intr^6\}~. 
}
This lattice is then used to define a six dimensional torus as the quotient $T^6 = \Real^6/\gG$ defined by the periodicities: 
\equ{
X ~\sim~ X + e\, m~,
\quad 
m\in \Intr^6~,
}
where the basis vectors $e_i$ are combined to the vielbein $e=(e_i)$. To obtain an orbifold the torus is modded out by the action of a finite group $\mathbf{G}$. 
Such an orbifold group $\mathbf{G}$ is either generated by
\equ{
\text{twists:} \quad X ~\sim~ D_\rep{v}(\gth)\, X~, \quad \gth \neq \Id
} 
or by
\equ{
\text{roto-translations:} \quad 
X ~\sim~ D_\rep{v}(\gth)\, X + e\, q~, 
\quad 
q \in \Ratl^6~. 
}
Here $D_\rep{v}(\gth)$ defines a six dimensional matrix representation of the abstract finite group element $\gth\in\mathbf{G}$ and it is related to an integral matrix $\widehat{D}_\rep{v}(\gth) = e^{-1} D_\rep{v}(\gth) e$. As can be inferred from the defining relations above, roto-translations are twists accompanied by translations. In order that these translations $q$ are physical, i.e.\ cannot be absorbed by the torus lattice translations, $q\not\in\Intr^6$. 
The space group $\mathbf{S}$ combines the torus lattice and the orbifold group elements: 
\equ{
(\Id; e\, m) \in \mathbf{S}~, \quad m \in \Intr^6~
\quad\text{and}\quad  
(\gth; e\, q) \in \mathbf{S}~, \quad \gth\neq \Id~,\quad q \in\Ratl^6~. 
}
The finite point group $\mathbf{P}$ is a projection of the space group $\mathbf{S}$: 
\equ{
\mathbf{S} ~\ra~ \mathbf{P}:
\quad 
(\gth, e\, q)  \mapsto \gth~, 
}
hence, the point group retains only the twist information of the orbifold group $\mathbf{G}$. The group multiplication of the space group elements $g=(\gth; e\, q)$ and $g'=(\gth'; e\,q')$ is given by 
\equ{ 
g \cdot g' ~=~ (\gth; e\, q) \cdot (\gth'; e\, q') 
~=~ (\gth \gth'; D_\rep{v}(\gth)\, e\, q' + e\, q)~. 
}
Consequently, space group elements may not commute because their point group parts $\gth$ and $\gth'$ do not commute or because their action on the lattice does not: 
\equ{
\big( \widehat{D}_\rep{v}(\gth) - \Id\big)\, q' ~\neq~ \big( \widehat{D}_\rep{v}(\gth') - \Id\big)\, q~. 
}

The action $D_\rep{v}(\gth)$ associated to a given space group element $g=(\gth, e\, q)$ can be diagonalized in a complex coordinate basis as 
\equ{
D_\rep{v}(\gth)  ~=~  \pmtrx{ 
e^{2\pi i\, v_g^1} & 0 & 0 \\[1ex]  
0 & e^{2\pi i\, v_g^2} & 0 \\[1ex]  
0 & 0 & e^{2\pi i\, v_g^3} }
}
in terms of a local twist vector $v_g=\big( 0, v_g^1,v_g^2,v_g^3\big)$. The characterization {\em local} here emphasizes that this diagonalization procedure has to be applied to each space group element $g$ separately. We reserve the term {\em global} to emphasize properties that hold for all space group elements. Unless the corresponding point group elements commute, the local twist vectors of different space group elements are defined with respect to different bases. For string theory we should be able to define spinors on the six dimensional torus. Assuming for now that they exist, the action of a space group element on an eight-component spinor is given by 
\equ{
D_\rep{s}(\gth) ~=~ 
e^{2\pi i\, v_g^1\, \frac {\gs_3}2} \otimes e^{2\pi i \, v_g^2\, \frac {\gs_3}2} \otimes e^{2\pi i\, v_g^3\, \frac {\gs_3}2}~. 
}
The expressions for $D_\rep{s}(\gth)$ and $D_\rep{v}(\gth)$ show that $\Spin{6} = \SU{4}$ is the double cover of $\SO{6}$: 
Both $D_\rep{s}(\gth)$ and $-D_\rep{s}(\gth)$ are associated to the same element $D_\rep{v}(\gth)$, because $D_\rep{v}(\gth)$ is inert under $v_g^a \mapsto v_g^a+1$, while $D_\rep{s}(\gth)$ changes sign.

The properties of the spinorial representation $D_\rep{s}(\gth)$ decide on how many supersymmetries a given space group element preserves, i.e.\ how many Killing spinors it admits. A space group element $g=(\gth; e\, q)$ admits a Killing spinor $\gPs_\text{inv.}$, if  
\equ{
D_\rep{s}(\gth)\, \gPs_\text{inv.} ~=~ \gPs_\text{inv.}
}
has non-trivial solutions $\gPs_\text{inv.}\neq 0$. 
Given that the possible eigenvalues of $D_\rep{s}(\gth)$ 
are $\exp(\pm 2\pi i\, \widetilde{v}_g^a)$, $a=0,1,2,3$, where
\equ{ 
\widetilde{v}_g ~=~
\frac 12 \pmtrx{
\phantom{-}v_g^1 + v_g^2 +  v_g^3 \\[1ex] 
-v_g^1 + v_g^2 +  v_g^3 \\[1ex] 
\phantom{-}v_g^1 - v_g^2 +  v_g^3 \\[1ex] 
\phantom{-}v_g^1 + v_g^2 -  v_g^3 
}
}
for a space group element $g=(\gth; e\, q)$ to admit at least one Killing spinor, at least one of the entries of $\widetilde{v}_g$ needs to vanish modulo integers. 
Consequently, $-\Id\in\Spin{6}$ breaks all supersymmetries, since the corresponding local twist vector would be $v_g=(0,\frac 12, \frac 12,\frac 12)$ so that none of the components of $\widetilde{v}_g = (\frac 34, \frac 14,\frac 14,\frac 14)$ vanish modulo integers. 
Since, the two choices for the spinor embedding of a space group element $g=(\gth,e\,q)$ differ by $-\Id$, at most one choice of $D_\rep{s}(\gth)$ admits some Killing spinors.

\section{Vanishing of the one-loop partition function}

As emphasized in the introduction the cosmological constant is a finite calculable quantity in string theory. In heterotic string theory the four dimensional one-loop cosmological constant $\gL_\text{4D}$ is computed as the integral of the full partition function $\mathcal{Z}_\text{full}$: 
\equ{ 
\gL_\text{4D} ~\sim~ 
\int_\cF \frac{\d^2\gt}{\gt_2^2}\, \mathcal{Z}_\text{full}(\gt,\bgt)~. 
\qquad \qquad 
\raisebox{-5ex}{\includegraphics[width = 3 cm]{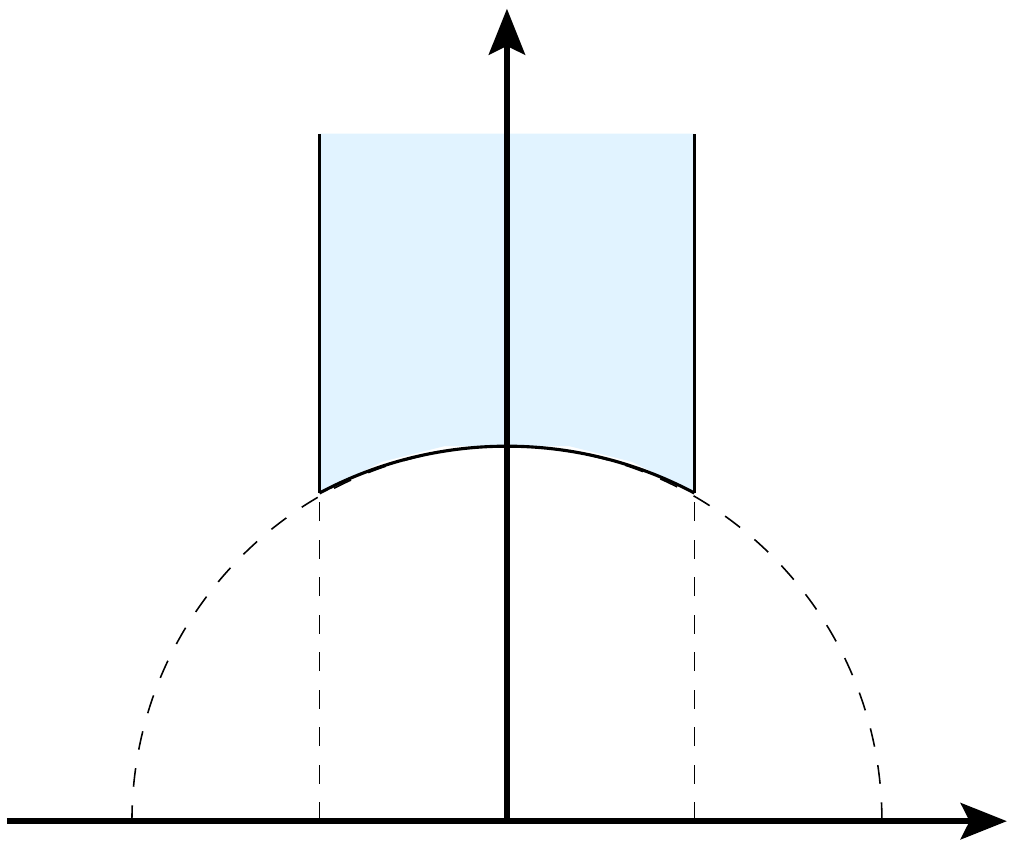}} 
}
The standard choice for the fundamental domain $\cF$ of the one-loop Teichmueller parameter $\tau$ is depicted by the blue shaded region in the picture next to the expression for the one-loop cosmological constant. Given that the integral over the fundamental domain $\cF$ is very complicated, we asked: 
Can we construct non-supersymmetric heterotic orbifolds which have a vanishing one-loop partition function? 

To investigate this possibility, let us consider what the full partition function consists of. The full partition function 
\(
\mathcal{Z}_\text{full} = \mathcal{Z}_\text{4D\,Mink.} \, \mathcal{Z}_\text{6D\,int.}
\)
can be divided into a four dimensional Minkowskian non-compact partition function $\mathcal{Z}_\text{4D Mink.}$, which never vanishes, and a six dimensional internal orbifold partition function \cite{Dixon:1986jc}
\equ{ \label{IntTwistedPI}
\mathcal{Z}_\text{4D\,Mink.}  ~=~ \frac 1{\gt_2} \Big| \frac1{\get^{2}} \Big|^2\neq 0
\quad\text{and}\quad 
\mathcal{Z}_\text{6D\,int.} ~=~ \frac 1{|\mathbf{P}|} \sum_{[g,h]=0} 
\mathcal{Z}_X\brkt{g}{h} \, 
\overline{\mathcal{Z}_\gps\brkt{g}{h}}\, 
\mathcal{Z}_Y\brkt{g}{h}~, 
}
respectively. The latter itself can be divided into three parts that are associated with the 6D internal coordinate fields $X$, their worldsheet superpartners described by the right-moving fermions $\gps$, and the 16D left-moving gauge degrees of freedom $Y$. The sum indicated here is over all commuting space group element $g,h\in\mathbf{S}$ normalized by the finite number of point group elements $|\mathbf{P}|$.

Contrary the other parts of the internal partition function, the right-moving fermionic partition function 
\equ{ 
\mathcal{Z}_\gps\brkt{g}{h}(\tau) = 
\frac 12\, e^{-\pi i\, v_g^T(v_h-e_4)}\,  \sum_{s,s'=0}^{1}\, (-)^{s's}\, 
e^{-2\pi i\, \sfrac{s'}2 e_4^T v_g} \, 
 \frac{\gth_4\brkt{\sfrac {1-s}{2} e_4- v_g}{\sfrac {1-s'}{2} e_4-v_h}}{\get^4}~, 
}
with $e_4=(1,1,1,1)$ and the product of four theta functions 
\equ{
\gth_4\brkt{\ga}{\ga'} = 
 \sum_{n\in\Intr^4} e^{2\pi i \big\{ \frac \gt2 (n+\ga)^2 + (n+\ga)^T\ga' \big\}}~, 
}
may vanish under certain circumstances. Using Riemann identities one can show that: 
\equ{ 
\mathcal{Z}_\gps\brkt{g}{h} =0 \quad \Leftrightarrow \quad
\arry{l}{
\text{$g, h \in \mathbf{S}$ share at least one Killing spinor.}
}
}
This confirms the well-known result that all supersymmetric orbifolds have vanishing partition functions. Also any non-supersymmetric toroidal orbifold for which
\enums{
\item[i.]  
a Killing spinor exists \textit{locally} in every commuting $(g,h)$-sector;
\item[ii.] 
but none \textit{globally,} 
}
will have a vanishing one-loop partition function. The second condition excludes globally supersymmetric orbifolds. 

Do such orbifolds exist? On first sight one might be inclined to say yes, since there are orbifold examples with different local and global supersymmetry breaking patterns: 
Consider for example the space group $\mathbf{S}$ of the DW(0--2) $\Intr_2\times\Intr_2$ orbifold \cite{Donagi:2004ht,Donagi:2008xy}
generated by the two orbifold elements: 
$g_\gth = \big(\theta, 0\big)$, $g_\go =\big(\omega, \sfrac 12\, e_5\big)$ and 
the torus translations $g_i =\big(1, e_i\big)$. 
The two-tori fixed by these two orbifold elements are depicted schematically below: 
\[
 \centerline{\scalebox{0.7}{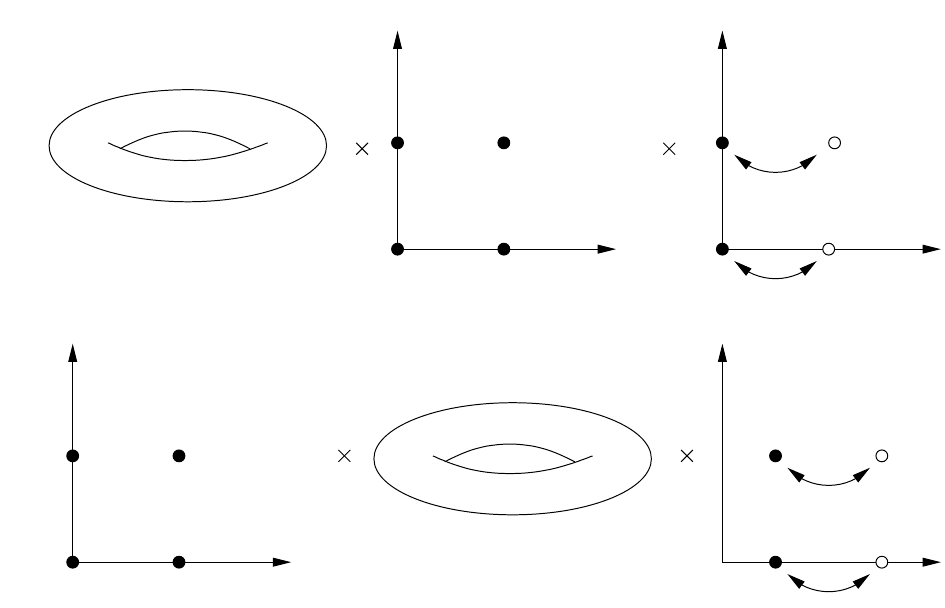}}
\]
Each of these two-tori preserve $\cN=2$ supersymmetry from a four dimensional perspective. However, the fixed tori of $g_\gth$ and $g_\go$ perserve different $\cN=2$ supersymmetries such that this orbifold only preserves $\cN=1$ globally. 
Yet, nowhere in the orbifold geometry supersymmetry is broken to $\cN=1$ locally, since the fixed two-tori of $g_\gth$ and $g_\go$ are shifted with respect to each other by $\frac 14\, e_5$ and therefore never intersect.

\section{Classification of toroidal orbifolds}

In light of this example it seems very reasonable to assume, that also orbifolds exist, for which locally always some amount of Killing spinors are preserved, but none globally. To investigate this possibility, we make use of the fact that all six dimensional toroidal orbifolds have been classified \cite{Opgenorth:1998ab,Plesken:2000ab}. This classification has three levels: 
There are 7,103 $\Ratl$-classes enumerating all the inequivalent point groups $\mathbf{P}$. Each of these point groups may act on various inequivalent lattices $\gG$ leading to 85,308 $\Intr$-classes. Finally, there are 28,927,915 affine-classes, corresponding to all the inequivalent space groups $\mathbf{S}$ (differing by their roto-translations) that can act on these lattices. 
Among these close to 29 million toroidal orbifolds only 520 preserve at least $\cN=1$ supersymmetry with 60 distinct underlying $\Ratl$-classes \cite{Fischer:2012qj,Fischer:2013qza}. 

This orbifold classification can be used to identify the $\Ratl$-classes of the toroidal orbifolds, that admit Killing spinors in all sectors locally, but none globally \cite{GrootNibbelink:2017luf}: 
\begin{center}
\renewcommand{\arraystretch}{1.2}
\begin{tabular}{| r | p{11cm} |}
\hline
{\bf \# $\Ratl$-classes} & {\bf Restriction} \\
\hline\hline 
7,103~~ & All inequivalent geometrical point groups $\mathbf{P} \subset \text{O}(6)$ \\
1,616~~ & Orientable geometrical point groups $\mathbf{P} \subset \SO{6}$ \\
106~~  & No element from $\mathbf{P}$ rotates in a two-dimensional plane only \\
63~~   & Each element $\gth \in \mathbf{P}$ admits a choice with some local Killing spinors \\
60~~   & Geometrical point group compatible with some global Killing spinors \\
\hline
\end{tabular}
\renewcommand{\arraystretch}{1}
\end{center} 
This table starts from all 7,103 inequivalent geometrical point groups. Out of those only 1,616 preserve orientation which is mandatory in order to admit spinors on the geometry. The next cut is made by vetoing any point group element that rotates in a single two-dimensional plane only, since such rotations necessarily break any supersymmetry in at least one orbifold sector. Out of the 106 possible $\Ratl$-classes only 63 are such, that each point group element individually admits some Killing spinors. 

Since most of them correspond to the 60 $\Ratl$-classes of the globally supersymmetric orbifolds, this leaves three candidate $\Ratl$-classes for toroidal orbifolds with the desired properties: 
\[
\renewcommand{\arraystretch}{1.4}
\begin{array}{|c|c|c|c|c|}
\hline
{\textsc{carat}}\textbf{-}\textbf{index} &\textbf{Point group} & \textbf{Generator relations}                   & \textbf{Order} & \textbf{Local twist vectors} 
\\\hline\hline
 3375 & \text{Dic}_3 =\Intr_3 \rtimes \Intr_4  & \gth_1^4 = \gth_2^3 = \Id~,        & 12 & \big( \frac{1}{4}, \frac{1}{4},-\frac{1}{2}\big)
 \\ 
&   & \gth_2\,\gth_1\,\gth_2=\gth_1          &  &  \big( \frac{1}{3},-\frac{1}{3}, 0\big) 
 \\ \hline
 5751 & Q_8                                     & \gth_1^4 = \Id,~\gth_1^2 = \gth_2^2~, &  8 & \big( \frac{1}{4}, \frac{1}{4},-\frac{1}{2}\big)
 \\
 &                                 & \gth_1\,\gth_2\,\gth_1=\gth_2 &  & \big( \frac{1}{4},-\frac{1}{4}, 0\big) 
 \\ \hline 
 6737 & \text{SL}(2,3)                          & \gth_1^3 = \gth_2^4 = \Id~, & 24 & \big( \frac{1}{3}, \frac{1}{3},-\frac{2}{3}\big)
\\
&                        &  (\gth_2\, \gth_1)^2 = \gth_1^2\,\gth_2 & & \big( \frac{1}{4},-\frac{1}{4}, 0\big) 
 \\ \hline
\end{array}
\renewcommand{\arraystretch}{1}
\] 
The first column gives their \textsc{carat}-indices to identify the relevant $\Ratl$-classes. The second column provides their abstract point groups; their defining relations are given in the next column and their order is tabulated after that. The final column states the local twist vectors, obtained in two different bases. As can be easily confirmed, they indeed separately preserve some amount of supersymmetry. 

For these three candidate $\Ratl$-classes all possible inequivalent spinor representations of their point groups were constructed. In all cases there is at least one point group element that does not preserve any Killing spinor! This leads to the conclusion, that there does not exist any non-supersymmetric orbifold for which all point group elements separately preserve some Killing spinors \cite{GrootNibbelink:2017luf}.

\section{Nonexistence proof of locally but not globally supersymmetric orbifolds} 

The last paragraph of the previous section sketched an argument for the nonexistence of locally but not globally supersymmetric orbifolds by explicit construction of all spin representations associated to all (candidate) point groups $\mathbf{P}$. To confirm this result, we also made use of representation theory of abstract finite groups \cite{GrootNibbelink:2017luf}. In the following we provide some of the necessary ingredients of this analysis.

First of all, the 7,103 $\Ratl$-classes of 6D orbifolds provided by \textsc{carat} correspond to only 1,594 different abstract point groups. The reason for this is, that for a given abstract group there may exist several inequivalent representations as integral $6\times 6$-matrices. In light of the discussion above, the following representations of an abstract point group $\mathbf{P}$ are relevant: 
\[
\renewcommand{\arraystretch}{1.4}
\arry{|c|c||c|c|}{ \hline 
\multicolumn{2}{|c||}{\text{\bf Geometrical point group}} & 
\multicolumn{2}{c|}{\text{\bf Abstract point group}} 
\\\hline 
\text{Name} & \text{Matrix repr.}  & \text{Repr.} & \text{Character} 
\\\hline\hline 
\text{Trivial} & 1 & \rep{1} & \gch_\rep{1}=1 
\\\hline 
\text{Spinor} & D_\rep{s}(\gth) & \rep{4} & \gch_\rep{4}  
\\\hline 
\text{Vector} & D_\rep{v}(\gth)  & \rep{6} = [\rep{4}]_2 & \gch_\rep{6} 
\\\hline 
}
\renewcommand{\arraystretch}{1}
\]
To ensure that the four-dimensional spinor representation $\rep{4}$ can be thought of as the double cover of the six-dimensional vector representation $\rep{6}$, the former is assumed to be obtained as the two times anti-symmetrized $\rep{4}$ denoted by $[\rep{4}]_2$.

Now, consider any $\rep{4}$-representations of $\mathbf{P}$ (in particular, neither $\rep{4}$ nor $\rep{6}$ need to be irreducible) that fulfils the following three properties:   
\enums{
\item 
The $\rep{4}$ lies inside $\SU{4}=\Spin{6}$: 
On all conjugacy classes of $\mathbf{P}$:  $\gch_{[\rep{4}]_4} = 1$. 
\item 
The $\rep{6}=[\rep{4}]_2$ is isomorphic to a $\Ratl$-class:
There should be an integral $6\times 6$-matrix representation $\widehat{D}_\rep{v}$ within the \textsc{carat} $\Ratl$-classes such that $\gch_\rep{6} = \gch_\rep{v} = \Tr \widehat{D}_\rep{v}$. 
\item 
The $\rep{4}$ does not contain the trivial singlet representation. 
}
Any $\mathbf{G}$-invariant Killing spinor satisfies:
$D_\rep{4}(\gth)\, \Psi_\text{inv.} = \Psi_\text{inv.}$ for all $\mathbf{G} \subset \mathbf{P}$. 
Consequently, the projector on the $\mathbf{G}$-invariant subspace reads: 
\equ{
\mathcal{P}^{\mathbf{G}} = \frac{1}{|\mathbf{G}|}\sum_{\gth' \,\in\, \mathbf{G}} D_\rep{4}(\gth')
}
Hence, the number of $\mathbf{G}$-invariant Killing spinors is counted by: 
\equ{
\mathcal{N}^{\mathbf{G}} 
= \text{Tr}\big( \mathcal{P}^{\mathbf{G}}\big) 
= \frac{1}{|\mathbf{G}|} \sum_{\gth' \,\in\, \mathbf{G}} \text{Tr}\left(D_\rep{4}(\gth')\right) 
= \frac{1}{|\mathbf{G}|} \sum_{\gth' \,\in\, \mathbf{G}} \chi_\rep{4}(\gth') 
= \langle \chi_\rep{4}, \chi_\rep{1} \rangle_{\mathbf{G}} 
= n_\rep{1}^\mathbf{G}
} 
The number of $\mathbf{G}$-invariant Killing spinors equals the number of trivial singlets $n_\rep{1}^\mathbf{G}$ in the branching of a $\rep{4}$-representation of $\mathbf{P}$ into irreducible representations of $\mathbf{G}$.  
In particular, the number of {\em local} Killing spinors preserved by $\gth$ is given:
$\mathcal{N}^{\langle \gth\rangle} = n_\rep{1}^{\langle \gth\rangle}$, 
since any element $\gth \in \mathbf{P}$ of order $N_\gth$ generates a $\langle\gth\rangle \cong \Intr_{N_\gth} \subset \mathbf{P}$. 
Similarly, the number of {\em global} Killing spinors is given by: $\mathcal{N} = n_\rep{1}^{\mathbf{P}}$, 
e.g.\ how many trivial singlets the $\rep{4}$-representation contains. 
In other words, it would be possible to obtain a locally but not globally supersymmetric orbifold, if there exists a point group $\mathbf{P}$ such that for all $\gth \in \mathbf{P}$: $n_\rep{1}^{\langle\gth\rangle} > 0$, 
while $n_\rep{1}^\mathbf{P} = 0$. This explains the condition (iii) mentioned above. 

For each of the 1,594 different abstract groups $\mathbf{P}$ we considered all faithful (but in general reducible) $\rep{4}$-representations and required, that they satisfy the three conditions (i) -- (iii) stated above. By constructing all $\Intr_{N} \subset \mathbf{P}$ subgroups we showed, that for each remaining $\rep{4}$, there is at least one cyclic subgroup, for which the $\rep{4}$ does not contain the trivial $\Intr_{N}$-singlet representation. 
This implies that for all non-supersymmetric six dimensional toroidal orbifolds, there is always a sector without any local Killing spinor. Hence there do not exist any locally but not globally supersymmetric toroidal orbifolds.

\section*{Acknowledgements}
I would like to thank the organizers of the conference DISCRETE 2018 in Vienna and the convener Alon Faraggi, in particular, to give me the opportunity to speak at this conference. 
In addition, I would like to thank Johanna Knapp for the kind hospitality at the mathematical physics group at Vienna University. 
But most of all, I would like to express my gratitude to my collaborators Orestis Loukas, Andreas M\"utter, Erik Parr and Patrick Vaudrevange for the project \cite{GrootNibbelink:2017luf} on which this talk was based. 
Finally, I would like to thank Paul Fogarty, Orestis Loukas and Patrick Vaudrevange for proofreading the manuscript.

\end{document}

%% file: texdefs.tex
\renewcommand{\d}{\mathrm{d}}

\DeclareMathSymbol{\mg}{\mathrel}{symbols}{"1D}

\newcommand{\ga}{\alpha}

\newcommand{\gth}{\theta}

\newcommand{\gs}{\sigma}
\newcommand{\gt}{\tau}
\newcommand{\go}{\omega}

\newcommand{\gps}{\psi}
\newcommand{\get}{\eta}
\newcommand{\gch}{\chi}
\newcommand{\gG}{\Gamma}

\newcommand{\gL}{\Lambda}

\newcommand{\gPs}{\Psi}

\newcommand{\cF}{{\cal F}}

\newcommand{\cN}{{\cal N}}

\renewcommand{\Tr}{\mbox{Tr}}

\newcommand{\Id}{\mathbbm{1}}

\newcommand{\ra}{\rightarrow}

\newcommand{\beq}{\begin{equation}}
\newcommand{\eeq}{\end{equation}}
\newcommand{\barr}{\begin{array}}
\newcommand{\earr}{\end{array}}
\newcommand{\equ}[1]{\begin{gather*} #1 \end{gather*}}

\newcommand{\enums}[1]{\begin{enumerate} #1 \end{enumerate}}

\newcommand{\arry}[2]{\begin{array}{#1} #2 \end{array}}

\newcommand{\pmtrx}[1]{\begin{pmatrix} #1 \end{pmatrix}}

\newcommand{\sfrac}[2]{\mbox{$\frac{#1}{#2}$}}
\newcounter{oldcounter}

\newcommand{\bgt}{{\bar\tau}}

\newcommand{\Intr}{\mathbbm{Z}}

\newcommand{\Real}{\mathbbm{R}}
\newcommand{\Ratl}{\mathbbm{Q}}

\newcommand{\ba}[2]{\[\begin{array}{#2}\label{#1}}
\newcommand{\ea}{\end{array}\]}
\newcommand{\be}{\begin{equation}}
\newcommand{\ee}{\end{equation}}
\newcommand{\bea}{\begin{eqnarray}}
\newcommand{\eea}{\end{eqnarray}}

\newcommand{\SU}[1]{\mathrm{SU(#1)}}
\newcommand{\SO}[1]{\mathrm{SO(#1)}}

\newcommand{\Spin}[1]{\mathrm{Spin(#1)}}

\newcommand{\brkt}[2]{\bigl[ ^{#1}_{#2} \bigr]}

\newcommand{\rep}[1]{\mathbf{#1}}

%% file: FixedTori_DW0-2.pdf_t
\begin{picture}(0,0)%
\includegraphics{FixedTori_DW0-2.pdf}%
\end{picture}%
\setlength{\unitlength}{2486sp}%
\begingroup\makeatletter\ifx\SetFigFont\undefined%
\gdef\SetFigFont#1#2#3#4#5{%
  \reset@font\fontsize{#1}{#2pt}%
  \fontfamily{#3}\fontseries{#4}\fontshape{#5}%
  \selectfont}%
\fi\endgroup%
\begin{picture}(7185,4626)(1111,-4459)
\put(1486,-1726){\makebox(0,0)[lb]{\smash{{\SetFigFont{12}{14.4}{\rmdefault}{\mddefault}{\updefault}$g_\gth$-fixed two-tori}}}}
\put(4006,-4111){\makebox(0,0)[lb]{\smash{{\SetFigFont{12}{14.4}{\rmdefault}{\mddefault}{\updefault}$g_\go$-fixed two-tori}}}}
\put(5806,-1996){\makebox(0,0)[lb]{\smash{{\SetFigFont{7}{8.4}{\rmdefault}{\mddefault}{\updefault}$e_3$}}}}
\put(3601,-16){\makebox(0,0)[lb]{\smash{{\SetFigFont{7}{8.4}{\rmdefault}{\mddefault}{\updefault}$e_4$}}}}
\put(8281,-1996){\makebox(0,0)[lb]{\smash{{\SetFigFont{7}{8.4}{\rmdefault}{\mddefault}{\updefault}$e_5$}}}}
\put(6076,-16){\makebox(0,0)[lb]{\smash{{\SetFigFont{7}{8.4}{\rmdefault}{\mddefault}{\updefault}$e_6$}}}}
\put(3331,-4381){\makebox(0,0)[lb]{\smash{{\SetFigFont{7}{8.4}{\rmdefault}{\mddefault}{\updefault}$e_1$}}}}
\put(1126,-2401){\makebox(0,0)[lb]{\smash{{\SetFigFont{7}{8.4}{\rmdefault}{\mddefault}{\updefault}$e_2$}}}}
\put(8281,-4381){\makebox(0,0)[lb]{\smash{{\SetFigFont{7}{8.4}{\rmdefault}{\mddefault}{\updefault}$e_5$}}}}
\put(6076,-2401){\makebox(0,0)[lb]{\smash{{\SetFigFont{7}{8.4}{\rmdefault}{\mddefault}{\updefault}$e_6$}}}}
\end{picture}%